\documentclass[a4paper, 11pt]{article}
\usepackage{tikz} 
\usepackage{amsmath}
\usepackage{amssymb}
\usepackage{hyperref}
\usepackage{listings}
\usepackage{dsfont}
\usepackage{graphicx}
\usepackage{subfigure}
\usepackage{multirow}
\usepackage[round]{natbib} 
\usepackage{enumitem}
\usepackage{authblk}
\usepackage[a4paper]{geometry}
\usepackage{proof}
\usepackage[ruled,vlined]{algorithm2e}
\usepackage{bm}
\usepackage{tabu} 
\usepackage{multirow}
\usepackage{comment}
\usepackage{xspace}
\usepackage{nicefrac}
\usepackage{lipsum}
\usepackage{lineno}
\usepackage[toc,page]{appendix}
\usepackage{breqn}
\usetikzlibrary{positioning}

\newcommand*\justify{%
  \fontdimen2\font=0.4em
  \fontdimen3\font=0.2em
  \fontdimen4\font=0.1em
  \fontdimen7\font=0.1em
  \hyphenchar\font=`\-
}

\renewcommand{\texttt}[1]{%
  \begingroup
  \ttfamily
  \begingroup\lccode`~=`/\lowercase{\endgroup\def~}{/\discretionary{}{}{}}%
  \begingroup\lccode`~=`[\lowercase{\endgroup\def~}{[\discretionary{}{}{}}%
  \begingroup\lccode`~=`.\lowercase{\endgroup\def~}{.\discretionary{}{}{}}%
  \catcode`/=\active\catcode`[=\active\catcode`.=\active
  \justify\scantokens{#1\noexpand}%
  \endgroup
}

\title{Integrating opportunities and parametrized signatures \\ for improved mutational processes estimation \\ in extended sequence contexts} 
\author{Ragnhild Laursen$^{1,2,\dagger}$, Marta Pelizzola$^{1,\dagger}$, Lasse Maretty$^{2,3,4}$,  and Asger Hobolth$^1$}
\date{  %
    {\small $^1$Department of Mathematics, Aarhus University.  \\
    $^2$Department of Molecular Medicine, Aarhus University \\
    $^3$Bioinformatics Research Center, Aarhus University. \\
    $^4$Current address:  QIAGEN Digital Insights, QIAGEN Aarhus, Denmark.\\
    $\dagger$ These authors contributed equally. \\
    E-mails: ragnhild@clin.au.dk, marta@math.au.dk, lasse.maretty@protonmail.com, asger@math.au.dk} \\%
    [2ex]%
    \today
}

\begin{document}

\maketitle

\section*{Abstract}
Mutational signatures describe the pattern of mutations over the different mutation types. Each mutation type is determined by a base substitution and the flanking nucleotides to the left and right of that base substitution. Due to the widespread interest in mutational signatures, several efforts have been devoted to the development of methods for robust and stable signature estimation. Here, we combine various extensions of the standard framework to estimate mutational signatures. These extensions include  (a) incorporating opportunities to the analysis, (b) allowing for extended sequence contexts, (c) using the Negative Binomial model, and (d) parametrizing the signatures. We show that the combination of these four extensions gives very robust and reliable mutational signatures. In particular, we highlight the importance of including mutational opportunities and parametrizing the signatures when the mutation types describe an extended sequence context with two or three flanking nucleotides to each side of the base substitution. \\
\textbf{Keywords:} Cancer genomics, mutational signatures, Negative Binomial, non-negative matrix factorization, opportunities, parametrization, sequence context.


\section{Introduction}
\subsection{Mutational signatures in cancer genomics}
The collection of mutations observed in a cancer patient is often referred to as the mutational catalogue of the patient. In this context, mutations are usually defined by taking into account the base substitution and its right and left flanking nucleotides, i.e.\, using the single-base-substitution-96 (SBS-96) mutational context  \citep{alexandrov2013signatures}. Nowadays the genome of numerous cancer patients have been sequenced and their mutational catalogues are available in public repositories such as the Pan-Cancer Analysis of Whole Genomes (PCAWG) database \citep{Campbell2020}. 

\begin{figure}[htb!]
    \centering
    \includegraphics[width = \textwidth]{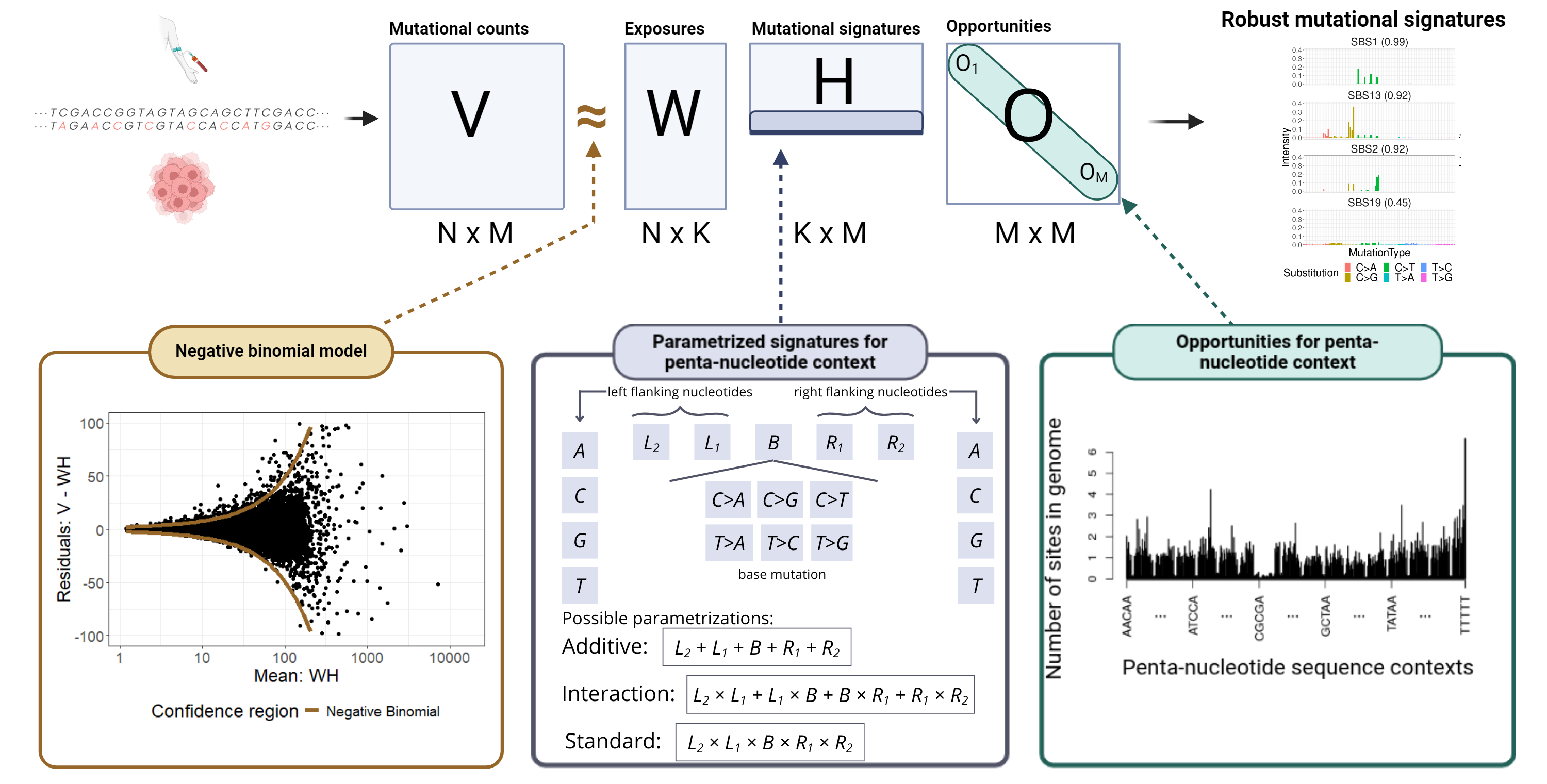}
    \caption{A graphical representation of the methods incorporated in our model to obtain robust mutational signatures. Top row: The mutational count data~$V$ is factorized into an exposure matrix~$W$, a mutational signature matrix~$H$, and a diagonal matrix of opportunities~$O$. Bottom row: Key ingredients of our model. The model assumes the Negative Binomial distribution for the counts (Negative binomial model), includes a parametrization of the signatures (Parametrized signatures for penta-nucleotide context), and takes into account the opportunity for each mutation type (Opportunities for penta-nucleotide context). }
    \label{fig:graphrep}
\end{figure}

The catalogue of mutations observed in a cancer tissue is generated by a combination of different mutational processes acting on the cancer cells. From the large collections of sequenced data from patients it is possible to obtain mutational signatures (top row in Figure~\ref{fig:graphrep}). These are probability vectors over the possible mutation types and each describes the pattern of a mutational process operative in the evolution of the cancer genomes. Cancer is mostly driven by few mutations in the genome, thus mutational signatures are highly important for distinguishing driver mutations from other processes taking place in the genome. A thorough understanding of mutational signatures and their associations with DNA repair deficiencies or exposures to certain agents can therefore play a key role for defining treatments for cancer patients \citep{Caruso2017} or developing prevention strategies \citep{Zhang2021}. Well known examples of mutational processes leaving distinctive signatures include aging \citep{Risques2018}, UV light \citep{Shibai2017} and tobacco smoking \citep{Alexandrov2016}. Other examples and aetiologies of different mutational signatures can be found in \cite{Tate2019}. 

Mutational signatures are usually identified using non-negative matrix factorization (NMF) \citep{alexandrov2013breastcancer, Lyu2020, Lal2021, Pelizzola2023}. NMF takes as input a matrix of mutational catalogues for different patients and factorizes it into the product of two non-negative matrices. In our framework, these are a matrix of mutational signatures and a matrix of weights representing the contribution of each signature to the total mutation counts of the different patients. Different approaches to estimate the signatures from mutational count data are reviewed in \cite{BaezOrtega2017} and \cite{Omichessan2019}. 
\subsection{Sequence context and opportunities}
The mutation rate depends on the sequence context around each site \citep{Lindberg2019, Dietlein2020, kmerpapa2022}. Thus, it is essential to include the sequence context information around the base substitution when defining mutation types for the inference of mutational signatures (see bottom right illustration in Figure~\ref{fig:graphrep}). The SBS-96 mutational context has been commonly used in the literature to define the mutation types for the mutational count matrix used as input in NMF. This definition can be extended by including two, three or more flanking nucleotides at each side of the base substitution to better capture the information carried out by the sequence context of each mutation. Extending the sequencing context in this way leads to more sparse data sets. 

Furthermore, sets of nucleotide sequences occur with very heterogeneous rates along the genome, which leaves different opportunities for the mutation types to occur. The opportunity of a mutation type describes the number of sites in the genome where that specific mutation type can occur. A higher opportunity for a specific mutation type gives a higher probability of observing a mutation from that type. 

The disparity in opportunities for different mutation types can be observed in the SBS-96 context, but becomes even more pronounced when looking at extended sequence contexts with two or three flanking nucleotides at each side. If the information contained in the opportunities is incorporated in the framework for estimating mutational signatures, the inferred signatures can be generalized regardless of the considered sequence composition (e.g.\, between whole genome and exome sequencing). Furthermore, by taking into account the sequence context in a genome, we obtain a more accurate estimation of the relative contribution of the different mutational processes. 
\subsection{Models for mutational count data}
Mutational opportunities have previously been used in \cite{Fischer2013} and \cite{Gori2018} in connection with the classical Poisson model and in the SBS-96 sequence context. Recent work, however, has shown that mutational count data is over-dispersed (see bottom left illustration in Figure~\ref{fig:graphrep}). As a result, the Negative Binomial model is better suited for modeling mutational count data than the Poisson model \citep{Lyu2020, Pelizzola2023}. Mutational opportunities have not yet been included in Negative Binomial NMF models to estimate mutational signatures, and their importance in extended sequence contexts still needs to be explored. Ignoring opportunities in NMF can cause an overestimation of the relative contribution of certain types of mutations compared to the overall mutational burden of the cancer genome. This can in turn lead to inaccurate conclusions about the underlying mutational processes. 
\subsection{Parametrized signatures}
Larger sequence contexts leads to an exponential increase of the number of parameters in the standard NMF parametrization. We include two alternatives to reduce the number of parameters in the model by parametrizing the mutational signatures (see bottom middle illustration in Figure~\ref{fig:graphrep}) as described in \cite{laursen2024}. This leads to more robust and interpretable signatures. We mainly investigate three types of parametrizations in this paper: the mono-nucleotide additive model of \cite{Shiraishi2015}, the di-nucleotide interaction model of \cite{laursen2024}, and the standard Poisson-NMF model of \cite{alexandrov2013signatures}.  
\subsection{The full model and overview of the paper}
Our final model thereby, for any sequence context, integrates opportunities and parametrized signatures into a Negative-Binomial-NMF framework. A graphical representation of our approach and contributions is shown in Figure~\ref{fig:graphrep}, and further elaborated in Section~\ref{sec:methods}. 

The paper is structured in the following way. Section \ref{sec:methods} starts by introducing the mathematical details of NMF for a general sequence context, and then Sections~\ref{subsec:NBNMF}, \ref{subsec:NMFopp} and \ref{subsec:NMFopppar} describe our extensions: Negative-Binomial distribution assumption, opportunities and parametrization. We derive new update rules for NMF where we include the mutational opportunities in the model directly. In Section \ref{sec:results}, we show that the combination of a Negative Binomial NMF model with the opportunities and parametrized signatures leads to robust signature estimation and generalizes well to unseen data. In order to show this, we consider two data sets from the PCAWG database \citep{Campbell2020}. Results on the breast cancer data set are discussed in Section \ref{subsec:breastres} and results on the liver cancer data set are summarized in Section \ref{subsec:liverres}. The paper ends with a discussion in Section~\ref{discussion}. 

The \verb+R+ implementation of our model and the code for reproducing our results are available at  \url{https://github.com/MartaPelizzola/ParOppR}.
\section{Methods} \label{sec:methods}
Consider a mutational count data set $V \in \mathbb{N}_0^{N \times M}$ where $N$ is the number of patients and $M$ is the number of mutation types. Mutational signatures are commonly derived by NMF \citep{Lee1999} which factorizes $V$ into the product of two non-negative matrices $W \in \mathbb{R}_+^{N \times K}$ and $H \in \mathbb{R}_+^{K \times M}$ such that 
\[V \approx WH.\]
In cancer genomics, the rows in $H$ represent the mutational signatures defined by probability vectors over the different mutation types, i.e.\, $H$ is normalized such that its rows sum to one. The matrix $W$ includes the exposures, where each row contains the exposure of each mutational signature for the corresponding patient. 

In the SBS-96 mutational context of length~3 \citep{alexandrov2013signatures} $M = 4 \cdot 6 \cdot 4=4^2 \cdot 6=96$, corresponding to the 6 base mutations when assuming strand symmetry times the 4 flanking nucleotides on each side.  As we want to go beyond this definition, we consider also  $M = 4^4 \cdot 6=1536$ and $M = 4^6\cdot 6=24576$, where sequence contexts of length $5$ and $7$ are used, respectively. We see that the number of mutation types increases exponentially with the sequence context. The classical model used when estimating mutational signatures has been proposed in \cite{Lee1999} and first applied to mutational count data in \cite{alexandrov2013signatures}, where the data is assumed to follow a Poisson distribution: 
\begin{align} \label{eq:poisson}
V_{nm} \sim  \text{Pois}\Big( (WH)_{nm} \Big).
\end{align} 
In order to estimate $W$ and $H$, we need to minimize the Kullback-Leibler divergence  
\begin{align} \label{eq:GKL}
     d_{\rm Po}(V||WH) = \sum_{n=1}^N \sum_{m=1}^M \left\{ V_{nm} \text{log} V_{nm} - V_{nm} \text{log}((W H)_{nm}) - V_{nm} + (W H)_{nm} \right\},
\end{align} 
which is equal to the negative Poisson log-likelihood up to an additive constant. 

Recently, \cite{Gori2018}, \cite{Lyu2020}, \cite{Vhringer2021} and \cite{Pelizzola2023} proposed an alternative model based on the Negative Binomial distribution to account for overdispersion in the mutational counts. We introduce this model in Section~\ref{subsec:NBNMF} and extend it in Section~ \ref{subsec:NMFopp} to include opportunities and in Section~\ref{subsec:NMFopppar} to include a parametrization of signatures.
\subsection{Negative Binomial NMF} \label{subsec:NBNMF}
This section summarizes the Negative Binomial NMF (NB-NMF) model with a patient specific dispersion coeffiecient. Here, 
\begin{equation} \label{eq:NB}
V_{nm} \sim \text{NB}\left( \alpha_n, \frac{(WH)_{nm}}{\alpha_n + (WH)_{nm}}\right), 
\end{equation}
where the mean and variance are given by
\begin{equation} \label{eq:meanvarNB}
    \mathbb{E}[V_{nm}] = (WH)_{nm} \quad \text{  and   } \quad  \mathbb{V}[V_{nm}] = (WH)_{nm} \left(1 + \frac{(WH)_{nm}}{\alpha_n}\right).
\end{equation}
The paramter $\alpha_n$ is the dispersion coefficient for each patient. Small values of $\alpha_n$ correspond to high dispersion in the data. On the contrary, when $\alpha_n \rightarrow \infty$, the Negative Binomial model in Equation \eqref{eq:NB} converges to the more commonly used Poisson model in Equation \eqref{eq:poisson}.

\cite{Pelizzola2023} showed that the negative log-likelihood of a Negative Binomial distribution is equal (up to an additive constant) to the following divergence
\begin{equation} \label{eq:divnb}
    \resizebox{0.9\textwidth}{!}{$
    d_{\rm NB}(V||WH) = \sum_{n=1}^N \sum_{m=1}^M \left\{ V_{nm} \log \left(\frac{V_{nm}}{ (WH)_{nm}}\right) -
    (\alpha_n + V_{nm}) \log \left(\frac{\alpha_n + V_{nm}}{\alpha_n + (WH)_{nm}} \right) \right\}. $}
\end{equation} 
Estimation of $W$ and $H$ is achieved using a Majorization-Minimization (MM) algorithm on this divergence measure (see e.g. \cite{Pelizzola2023}). 
Here, we propose to further extend the NB-NMF for extracting mutational signatures by including mutational opportunities in the model and parametrizing the mutational signatures especially for large mutational contexts. 
\subsection{NB-NMF with opportunities} \label{subsec:NMFopp}
\begin{figure}[h]
    \centering
    \includegraphics[width = 0.98\textwidth]{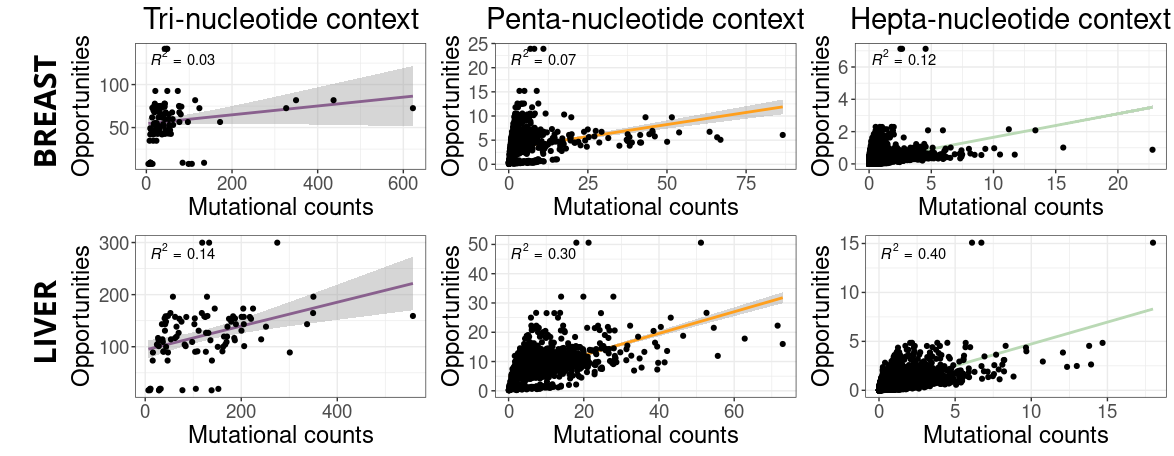}
    \caption{The average mutational count across patients for each mutation type is plotted against  the corresponding mutational opportunities for the breast (top panel) and liver cancer (bottom panel) patients in \cite{Campbell2020}. The mutational contexts with 1 (left), 2 (middle) and 3 (right) flanking nucleotides on each side are shown for each cancer type.}
    \label{fig:corrcountopp}
\end{figure}

Mutational opportunities correspond to the number of sites in the genome where a certain mutation type can occur. In Figure~\ref{fig:corrcountopp} the mutational opportunity for each mutation type is plotted against the mean number of mutations per patient for the corresponding mutation type. Patients from the breast and liver cancer data sets in \cite{Campbell2020} are used in this figure and each panel refers to one of the different mutational contexts. Figure \ref{fig:corrcountopp} demonstrates that the correlation between the observed counts and the opportunities increases when moving from the classical SBS-96 context to larger number of flanking nuelotides. The correlation for breast cancer increases from $R^2 = 0.03$ to 0.07 to 0.12 when the context size goes from three to five to seven nucleotides. The increased correlation is even more pronounced for liver cancer where $R^2$ goes from 0.14 to 0.3 to 0.4. Thus, accounting for the opportunities is needed when modeling the mutational counts and even more for the extended nucleotide contexts. Signature reconstruction from mutational count data with opportunities will provide a more accurate estimation of the relative contributions of different mutational processes. 

To incorporate opportunities into the framework we first model the mutational counts as coming from a binomial distribution as described in e.g. \cite{Weinhold2014} and \cite{Lochovsky2015}. If the mutation type is $m = A [T > C] G$, then $O_m$ is the number of times the triplet $ATG$ is observed in the genome and $p_{nm}$ is the probability that this specific mutation occurs at the sites $ATG$ in patient~$n$
\begin{equation} \label{eq:betadist}
    V_{nm} \sim \text{Bin}(O_{m},p_{nm}).
\end{equation}

Since $O_m$ is large and $p_{nm}$ is small, we can approximate the binomial model by the Poisson model where
\begin{align}
    \text{Bin}(O_m,p_{nm}) \simeq \text{Pois}(p_{nm} O_m) = \text{Pois}((WH)_{nm} O_m ). 
\end{align}
As $O_m$ is fixed we can write $p_{nm}$ as the factorization $(WH)_{nm}$ that needs to be estimated. This is an extension of the model presented in equation \eqref{eq:poisson}, where $(WH)_{nm}$ is replaced by $(WH)_{nm} O_m$. The Poisson model can be further extended to the Negative Binomial model by allowing additional dispersion in the model. 

The Negative Binomial model including opportunities has the following mean and variance: 
\[
  \mathbb{E}[V_{nm}] = (WH)_{nm}O_m \quad \text{  and   } \quad  \mathbb{V}[V_{nm}] = (WH)_{nm} O_m \left(1 + \frac{(WH)_{nm}O_m}{\alpha_n}\right),
\]

where~$\alpha_n$ is the dispersion coefficient for each patient and~$O_m$ is the opportunity for mutation type $m$. 

Under this model, the log-likelihood function is given by
\begin{equation*} 
    \resizebox{\textwidth}{!}{
    $\ell(W,H;V)  = 
    \sum_{n=1}^N \sum_{m=1}^M 
    \bigg\{ 
      \log { \binom{\alpha_n + V_{nm}  - 1}{\alpha_n}} + 
      V_{nm} \log \left( { \frac{(WH)_{nm}O_m}{\alpha_n + (WH)_{nm}O_m}}\right) + 
      \alpha_n \log \left( { 1 - \frac{(WH)_{nm}O_m}{\alpha_n + (WH)_{nm}O_m} } \right)
    \bigg\}$},
\end{equation*}  
and maximizing the log-likelihood is equivalent to minimizing the divergence measure 
\begin{equation} \label{eq:dNBO}
    \resizebox{0.93\textwidth}{!}{
    $d_{\rm NBO}(V||WH) =  
    \sum_{n=1}^N \sum_{m=1}^M V_{nm} \left\{ 
    \log \left(\frac{V_{nm}}{ (WH)_{nm}O_m}\right) - (\alpha_n + V_{nm}) \log \left(\frac{\alpha_n + V_{nm}}{\alpha_n + (WH)_{nm}O_m} \right) 
    \right\}.$}
\end{equation}
Notice, that this is almost equivalent to the divergence in \eqref{eq:divnb}, where $(WH)_{nm}$ is simply replaced by $(WH)_{nm} O_m$.
To derive the updates for~$W$ and~$H$ we use the MM-algorithm similar to~\cite{Gouvert2020} and~\cite{Pelizzola2023}.  

The resulting multiplicative updates for $H_{km}$ and $W_{nk}$ are as follows:
\begin{align}
   H_{km}^{t+1} = H_{km}^t \frac{\sum_{n=1}^N \frac{V_{nm}}{(WH^t)_{nm}O_m} W_{nk}}{\sum_{n=1}^N \frac{V_{nm} + \alpha_n}{(WH^t)_{nm}O_m + \alpha_n} W_{nk}}. \label{eq:updateH_methods}
\end{align}
and
\begin{align}
    W^{t+1}_{nk} = W^t_{nk} \frac{\sum_{m=1}^M \frac{V_{nm}}{(W^tH)_{nm} O_m} H_{km}}{\sum_{m=1}^M \frac{V_{nm} + \alpha_n}{(W^tH)_{nm}O_m + \alpha_n} H_{km}}.  \label{eq:updateW_methods}
\end{align}
The detailed derivations leading to these updates can be found in Appendix~\ref{app:A}. The procedure to estimate the patient-specific dispersion parameters $\bm{\alpha} = (\alpha_1, ..., \alpha_N)$ is equivalent to the one outlined in \cite{Pelizzola2023} with the inclusion of the opportunities. We obtain maximum likelihood estimates of $\bm{\alpha}$ based on the Negative Binomial likelihood using the Newton-Raphson method with the estimate of $WH$ from the Poisson NMF with opportunities. 
\subsection{NB-NMF with opportunities and parametrized signatures} \label{subsec:NMFopppar}
Opportunities are particularly important when working with extended context, where more flanking nucleotides to the base mutation is included as emphasized in Figure~\ref{fig:corrcountopp}. Here, the number of sites for different groups of nucleotides becomes increasingly heterogeneous. \cite{laursen2024} have described a framework to parametrize mutational signatures for an arbitrary large nucleotide context to avoid overfitting and obtain more stable signatures, which are also easier to intepret. The signatures are parametrized by the natural features of the mutation type, which considers the base substitution and each of the flaking nucleotides as separate variables. Here we are incorporating this framework into the NB-NMF model with opportunities.

We consider three main parametrized models for the signatures: the standard model from \cite{alexandrov2013signatures}, the additive model from \cite{Shiraishi2015} and the interaction model from \cite{laursen2024}. The models are illustrated in Figure~\ref{fig:graphrep} (middle figure in bottom row) when two flanking nucleotides are included. Let $f$ be the number of flanking nucleotides considered to each side of the base mutation. Using the notation from Figure~\ref{fig:graphrep}, the standard model is given by $L_f \times \cdots \times L_1 \times B \times R_1 \times \cdots \times R_f$. Thus, the number of parameters is $D = 6 \cdot 4^{2f}$, which means it increases exponentially with the number of flanking nucleotides. The additive model does not include any interaction term and can be written as $L_f + \cdots + L_1 + B + R_1 + \cdots + R_f$ and the number of parameters can be calculated as a linear function of the number of flanking nucleotides: $D = 6 + 2f \cdot (4-1) = 6 + 6f$. Lastly, in the interaction model the number of parameters is larger than for the additive model as interactions for neighboring nucleotides are considered. Though, the number of parameters still remain linear in the number of flanking nucleotides: $D = 6 \cdot 3 + 2f (3 \cdot 4) = 18 + 24f$. In this case, the parametrization with $f$ flanking nucleotides is given by $L_f \times L_{f-1} + \cdots + L_2 \times L_1 + L_1 \times B + B \times R_1 + R_1 \times R_2 + \cdots R_{f-1} \times R_f$.

All three models can be incorporated into the framework of \cite{laursen2024}, where each mutational signature is parametrized using the log-linear link function
\[  
  H_k = H_k(\beta_k) = \frac{1}{C_k} \exp(X_k \beta_k), 
\] 
where $C_k = \textbf{1}' \exp(X_k\beta_k)$ is a normalizing constant and $X_k$ is the design matrix of dimension $M \times D$, where $M$ is the number of mutation types and $D$ is the number of parameters in the parametrization. The entries in $X_k$ are fixed dependent on the specified parametrization of the mutation types. The values in $\beta_k$ are estimated using quasi-Poisson regression. More details on parametrizing mutational signatures are available in \cite{laursen2024}. 

Algorithm~\ref{alg:NMFopp} in Appendix~\ref{app:B} shows an MM-algorithm for the estimation of the factorization under the Negative Binomial model including both parametrization and opportunities. In each iteration we do the following: Update the mutational signatures under the NB-NMF model with opportunities in \eqref{eq:updateH_methods}, fit a log-linear multinomial model to the signatures (using the Poisson trick), update the signatures according to the estimated $\hat{\beta}_k$ and specified parametrization of the design matrix~$X_k$, and lastly update the exposure matrix~$W$ with the update rule from the NB-NMF model with opportunities in~\eqref{eq:updateW_methods}. 
\subsection{Model validation}
In our results, we express the cost in estimating the data in terms of the generalized Kullback-Leibler divergence (recall Equation \eqref{eq:GKL}) and we use the Bayesian Information Criterion (BIC)
\begin{align}\label{eq:BIC}
     \text{BIC} = 2d_{\rm NBO}(V||WH) + \ln(n_{obs}) n_{prm} 
\end{align}
to determine which models are more appropriate under the different scenarios. Here $d_{\rm NBO}(V||WH)$ is the log-likelihood from the Negative Binomial model (from Equation \eqref{eq:dNBO}), $n_{obs}$ is the number of observations and $n_{prm}$ the number of free parameters in the model and $n_{prm} = N \cdot K + K \cdot D$, where the value of $D$ depends on the chosen parametrization of the signatures. A model is thus penalized if it has a poor fit to the data (first term) and/or if it is highly complex, i.e. has a large number of parameters (second term). Models striking a good balance between the fit to the data and model complexity have small BIC values. We note that the the number of observations in this context can be set as the total number of counts (i.e.\, $N \cdot M$) or as the number of patients $N$, leading to an ambiguity in the definition due to the structure of the mutational matrix $V$. Here we set $n_{obs} = N \cdot M$.

\section{Results} \label{sec:results}
In this section we illustrate the results of the NB-NMF model with opportunities and parametrization for two data sets from the PCAWG database \citep{Campbell2020}. In this database 2782 patients from different cancer types are available and the mutational counts can be found at \url{https://www.synapse.org/#!Synapse:syn11726620}. We consider a data set of breast cancer patients in Section \ref{subsec:breastres} and one of liver cancer patients in Section \ref{subsec:liverres}. For both examples we show how the inclusion of opportunities and the parametrization improves the model for different number of signatures and parametrizations.

\subsection{Results on the breast cancer data set} \label{subsec:breastres}

In this section we extract mutational counts for the 120 breast cancer patients from the PCAWG database and use them for our analysis.  

\begin{figure}[h]
    \centering
    \begin{tikzpicture}
        \node[anchor=west] (bic) at (-2,0) {\includegraphics[height = 3.8cm]{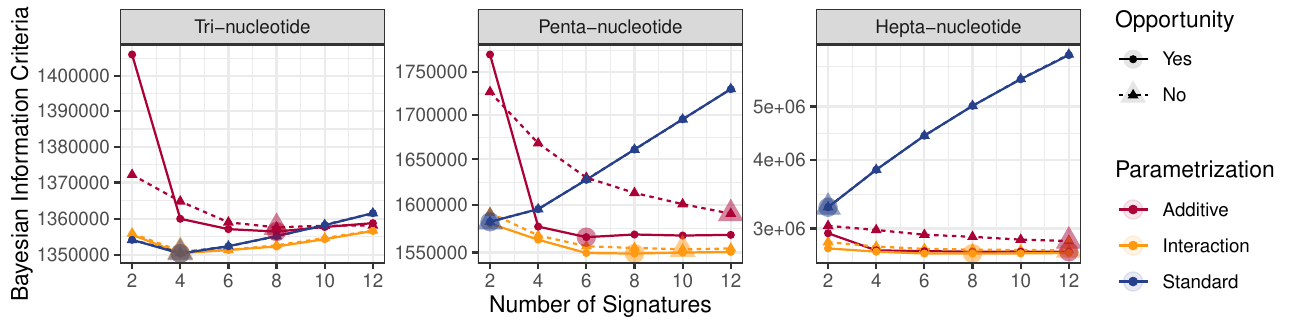} };
        \node[anchor=west] (alpha) at (0.5,-4.3) {\includegraphics[height = 3.8cm]{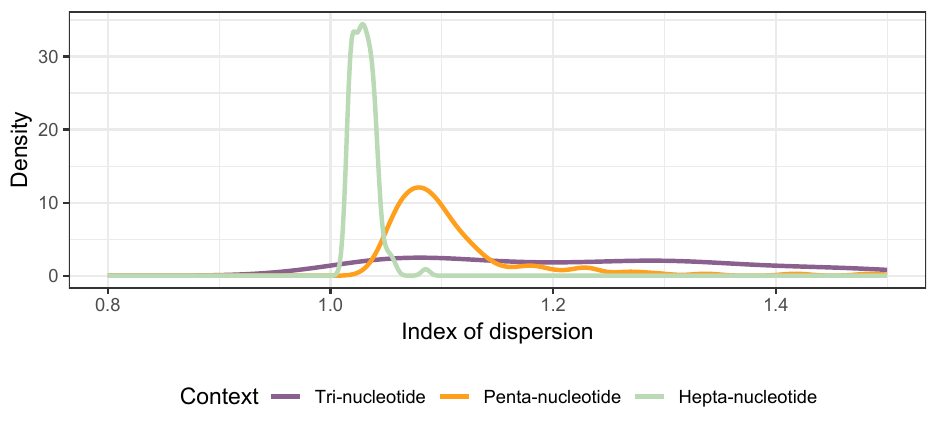}};
        \node[above left = -0.4cm of bic] {\textbf{A}};
        \node[above left = -0.4cm of alpha] {\textbf{B}}; 
    \end{tikzpicture}
    \caption{\textbf{A.} The influence of parametrizing and including opportunities on the breast cancer data set. The BIC in log-scale plotted against the number of signatures. The minimum is highlighted by a larger point.  \textbf{B.} Kernel density of the estimated dispersion index $\hat{D}_{nm}$ for the standard NB-NMF model with four signatures.}
    \label{fig:BRCAres}
\end{figure}
The results in Figure \ref{fig:BRCAres}A show the BIC against the number of signatures used for estimation. These results show that the additive \citep{Shiraishi2015} and the interaction models can achieve a good fit to data while keeping the number of parameters low. On the contrary, the number of parameters becomes large very quickly with the standard model when modeling data in a Penta-nucleotide or Hepta-nucleotide context.

\begin{figure}[h!]
    \centering
    \begin{tikzpicture}
        \node[anchor=west] (noopp) at (0,0) {\includegraphics[width = \textwidth]{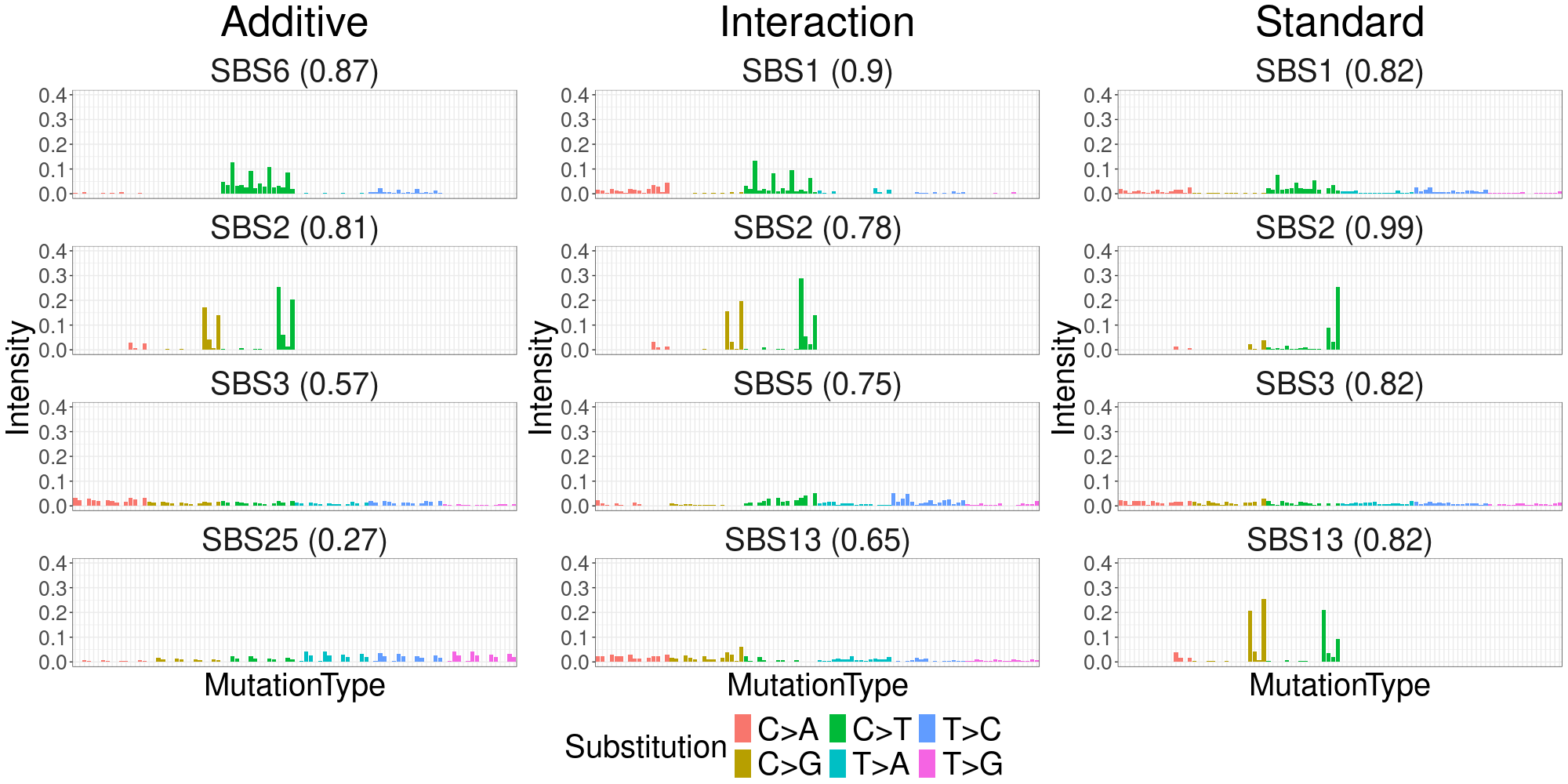} };
        \node[anchor=west,below=of noopp] (opp){\includegraphics[width = \textwidth]{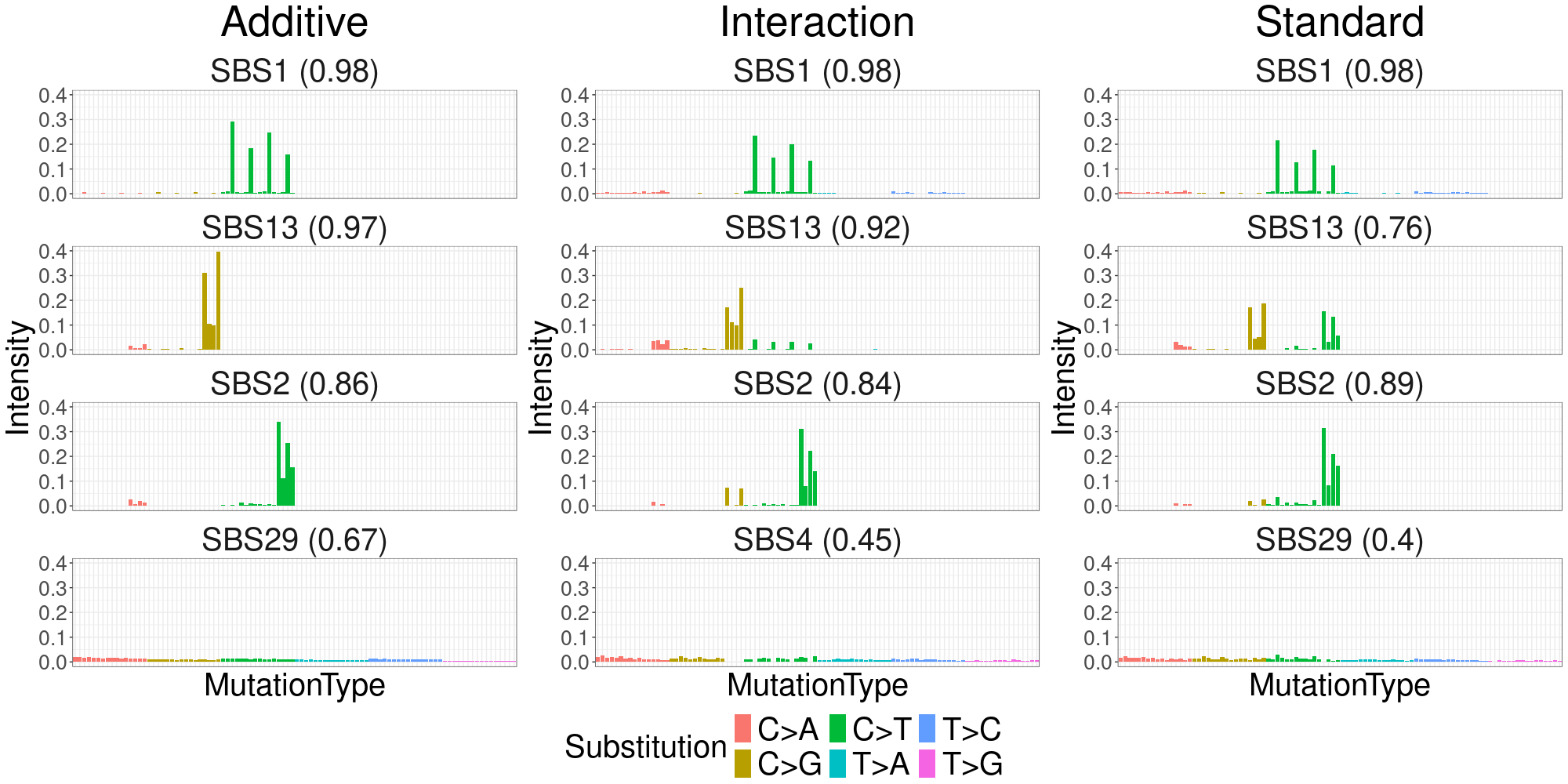} };
        \node[above left = 0.2cm of noopp,overlay, anchor = west] {\textbf{A - Without opportunity}};
        \node[above left= 0.2cm of opp,overlay, anchor = west] {\textbf{B - With opportunity}}; 
    \end{tikzpicture}

    \caption{The mutational signature for the models without opportunities \textbf{A} and with opportunities \textbf{B}. The number in the parenthesis is the cosine similarity between the constructed signatures from the additive, interaction and standard model to the COSMIC signatures for the SBS-96 context.}
    \label{fig:cosinesimBRCA}
\end{figure}

For the standard model, we see that the models with and without opportunities provide the same fit. This is due to equivalence in the model estimation besides the scaling of the opportunities. As the mean is simply scaled in the model with opportunities it is possible to obtain the same fit by scaling the mean parameter afterwards as follows: 
\[ (WH)_{nm}O_m = \sum_{k=1}^K W_{nk}H_{km}O_m = \sum_{k=1}^K W_{nk}(H_{km}O_m) = \sum_{k=1}^K W_{nk}\tilde{H}_{km} = (W\tilde{H})_{nm}.\]

For the parametrized models, including the opportunities (solid lines) returns lower BIC values and an improved fit. This is especially true for the Penta and Hepta-nucleotide contexts, where the difference between the cost of the models with and without opportunities becomes larger. In these contexts, we find that the best fit for the interaction model with opportunities corresponds to a model with consistently fewer signatures compared to the additive and standard model which is also expected \citep{Alexandrov2020}. These results also emphasize that the inclusion of the opportunities is essential in larger contexts as the BIC for the standard model shows that this model is too complex for this data.

From Equation \eqref{eq:meanvarNB} we get the dispersion index:
\begin{equation} \label{eq:dispindex}
    D_{nm} = \frac{\mathbb{V}[V_{nm}]}{\mathbb{E}[V_{nm}]} = 1 + \frac{(WH)_{nm}O_m}{\alpha_n}.
\end{equation}
The distribution of the estimated dispersion index $\hat{D}_{nm} = 1 + \frac{(\hat{W}\hat{H})_{nm}O_m}{\hat{\alpha}_n}$ is shown in Figure \ref{fig:BRCAres}B.
It is illustrated for the different sizes of nucleotide context from Figure \ref{fig:BRCAres}A in the scenario where four signatures is assumed. This dispersion index in \eqref{eq:dispindex} is exactly the value that scales up the variance in the Negative Binomial distribution compared to the Poisson distribution. In Figure \ref{fig:BRCAres}B we see that the estimated dispersion index is persistently larger than one, supporting the Negative Binomial as the most suitable model for mutational counts. Though, the dispersion index is decreasing as the context size increases. While this may be partially attributed to lower counts per mutation type in extended contexts, it likely reflects the unexplained heterogeneity of mutation rates in small contexts: by accounting for larger contexts, we capture specific mutation rates that were not visible within smaller context categories, which reduces the unexplained overdispersion.

In Figure \ref{fig:cosinesimBRCA} we have visualized the signatures for rank four for the three different parametrizations, with and without opportunities. These are compared to the results presented in \cite{Alexandrov2020} which also suggest that there are mainly 4 active signatures in breast cancer patients, namely SBS1, SBS2, SBS5, and SBS13. Furthermore, SBS3 and SBS18 are also moderately present in breast cancer patients. 

We identified the most similar signatures from the Catalogue Of Somatic Mutations In Cancer (COSMIC) version 3. Figure \ref{fig:cosinesimBRCA} shows the estimated signatures from the different models together with the cosine similarities to the most similar COSMIC signatures. The signatures from the COSMIC database are available at \url{https://cancer.sanger.ac.uk/cosmic}. 

We show results without the inclusion of opportunities in Figure  \ref{fig:cosinesimBRCA}A. Here, the interaction and standard models can reconstruct major signatures of breast cancer patients with high accuracy: the interaction model can reconstruct SBS1, SBS2, SBS5, and SBS13, whereas under the standard model SBS3 is reconstructed instead of SBS5. The average cosine similarity of the standard model ($0.86$) is larger than the one for the interaction model ($0.77$). The additive model is not able to estimate the four major signatures of breast cancer patients and has lower cosine similarity in this scenario. 

On the contrary, when opportunities are included the cosine similarity between the estimated signatures from both parametrized models with opportunities is much higher (Figure \ref{fig:cosinesimBRCA}B). All three models reconstruct signatures SBS1, SBS2, and SBS13 with high cosine similarity. Considering these three signatures, the average cosine similarity for the additive model is $0.94$, for the interaction model is $0.91$, and for the standard model is $0.88$. This shows that the parametrized models with opportunities can achieve very high accuracy in reconstructing signatures with a much simpler model. The last signature corresponds to a background signature. The cosine similarity formulation favors high peaks in the signatures and makes it hard to match background (flat) signatures to existing ones. Nonetheless, a background signature (namely SBS5) has been found in all cancer types \citep{Alexandrov2020}, and thus it is also expected in our application where indeed a background signature is successfully reconstructed by all models with opportunities (fourth row in Figure \ref{fig:cosinesimBRCA}B).

\subsubsection{Estimation with and without opportunities}
\begin{figure}[h]
    \centering
    \includegraphics[width = \textwidth]{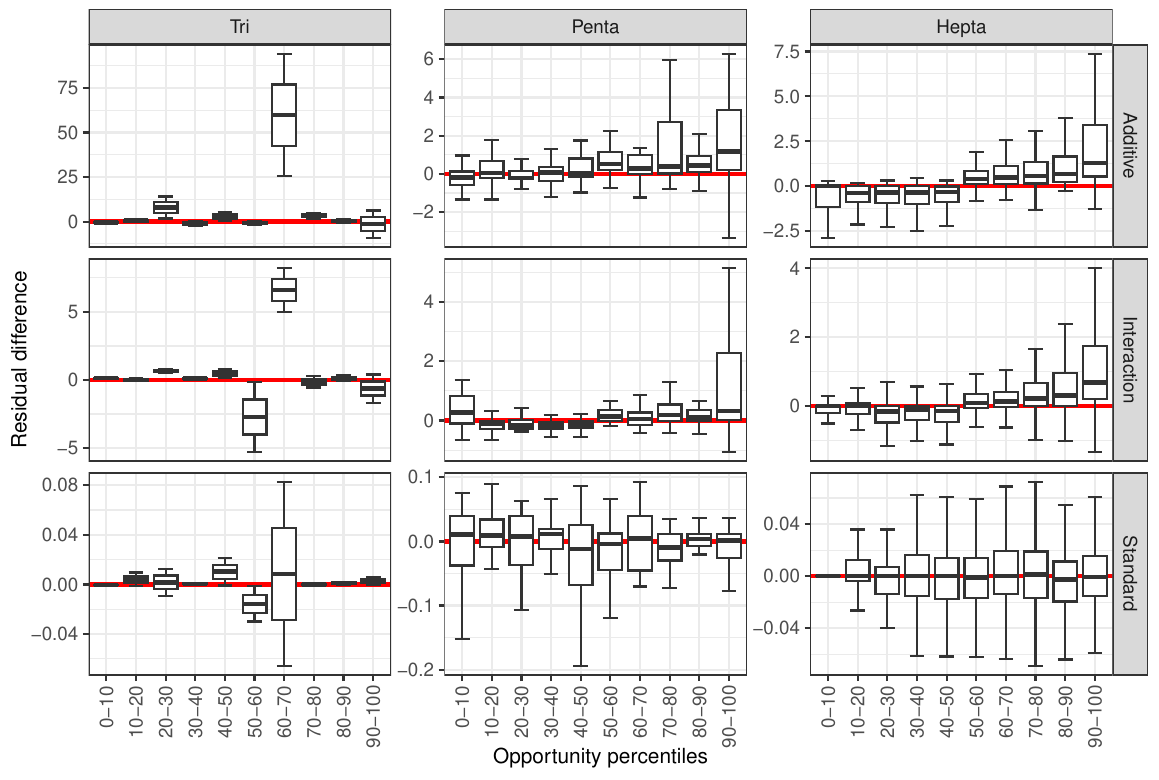}
    \caption{The difference in the standardized residuals in the model with and without opportunities, such that positive values means better prediction with opportunities. Each boxplot represent the results for mutation types with an opportunity between the specified percentile. The red line illustrate the zero line. }
    \label{fig:diff_opp}
\end{figure}

The results in Figure \ref{fig:BRCAres}A and \ref{fig:cosinesimBRCA} show how the opportunities improve the estimation of the data. To better understand under which conditions including the opportunities in the model offers most advantages we show the difference of the average standardized residuals with and without opportunities in Figure \ref{fig:diff_opp}. 
Each point here corresponds to the difference of the average standardized residuals with ($R_{nm}^{opp}$) and without ($R_{nm}$) opportunities for each mutation type $m$ as a function of its opportunity:
\[\bar{R}_{diff,m} = \Delta\bar{R}_{m} = \frac{1}{N} \sum_{n=1}^N (R_{nm}^{opp} - R_{nm}).\]
It shows that the model with opportunities gives better prediction of data points with a high opportunity when the context is defined including five or seven nucleotides. We also provide a visualization of the average standardized residuals $R_{nm}^{opp}$ and $R_{nm}$ as a function of the opportunity percentiles for the interaction model with three, five and seven flanking nucleotides in Appendix~\ref{app:C}. This shows that the residuals without opportunities have higher variability compared to those coming from the models with opportunities with the latter providing a better fit to the data.

As discussed in Section \ref{subsec:breastres} in relation to Figure \ref{fig:BRCAres}A there is no difference between the standard model with and without opportunities. The greater difference observed in the parametrized models with and without opportunities arises from the fact that these models involve parametrizing the signatures during estimation. Here the opportunities will change the signature vectors that are being parametrized in the estimation. In Figure \ref{fig:corrcountopp} it is shown that the opportunities are correlated with the mutational count and it is therefore expected to reduce variability in the signatures. As a result, it becomes more accurate to parametrize the signatures when opportunities are considered, leading to improved estimation in this case.

\subsubsection{Increase in accuracy of exposures estimation} \label{subsec:Waccuracy}
In this section we show that the inclusion of opportunities improves the accuracy of estimating the exposures. 
\begin{figure}[h!]
    \centering
    \includegraphics[width =0.9\textwidth]{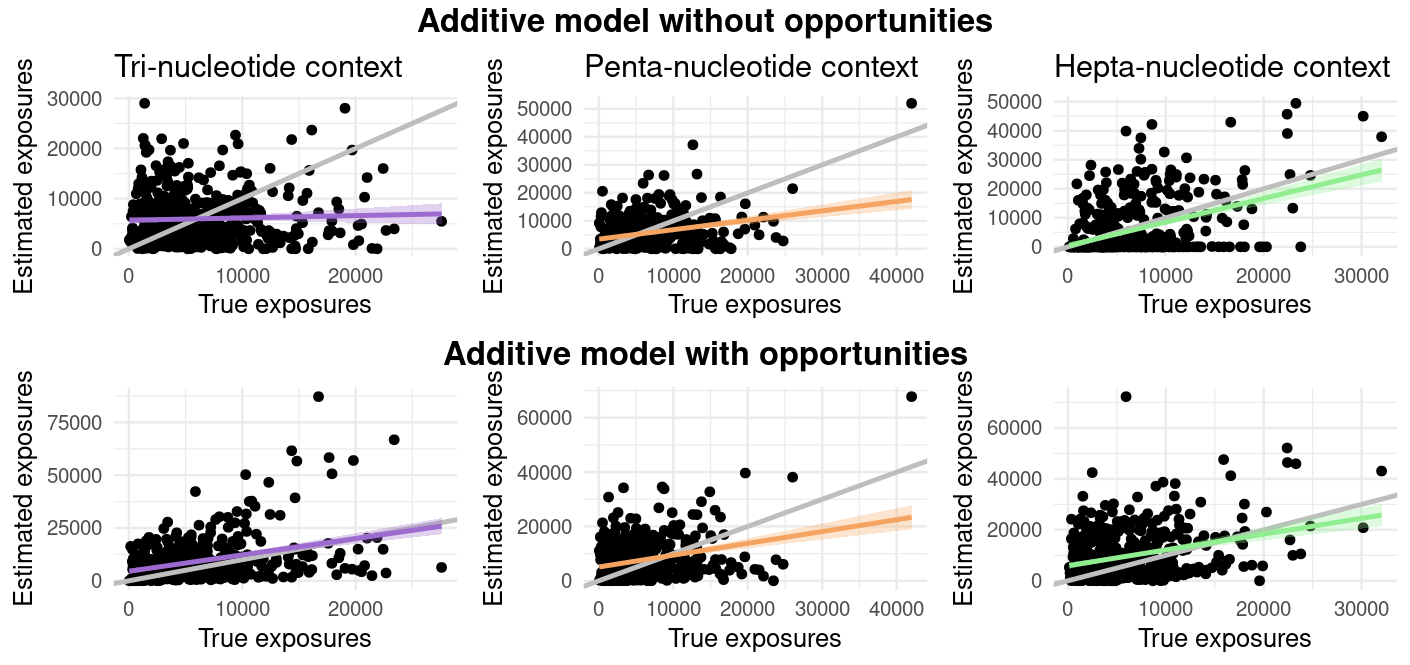}
    \includegraphics[width =0.9\textwidth]{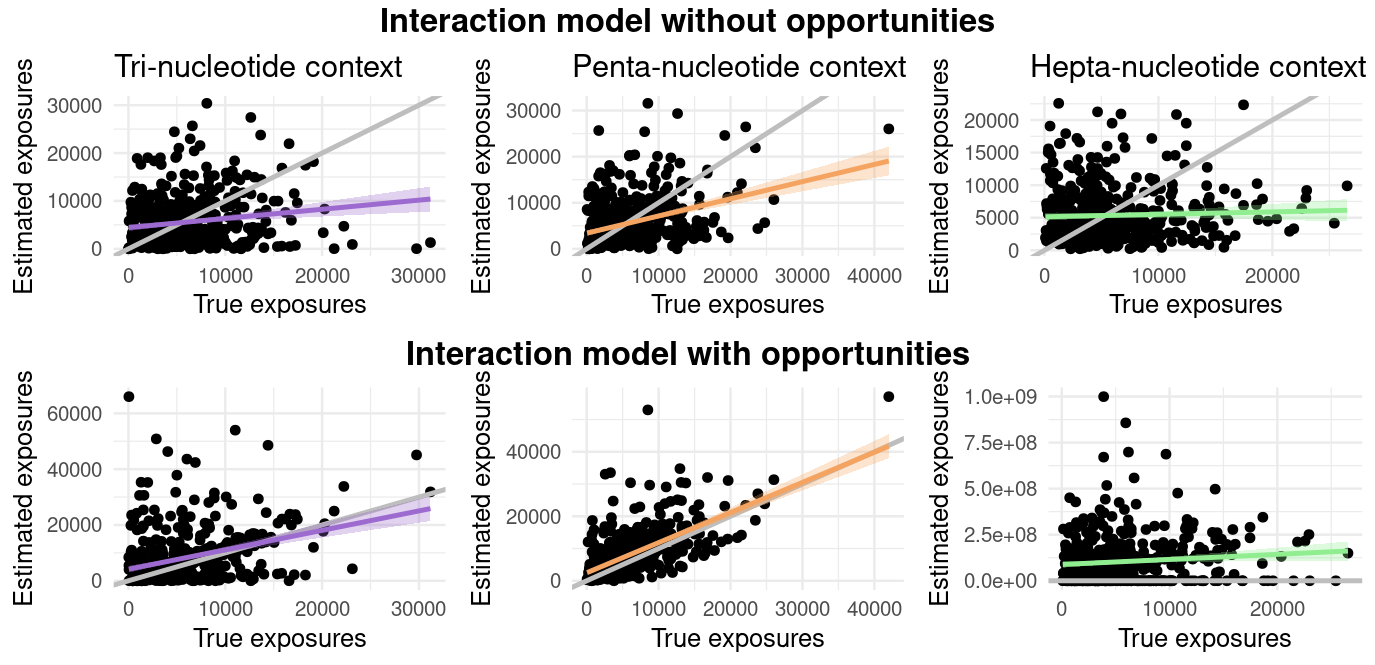}
    \caption{The estimated exposure for each patient and signature is plotted against the corresponding simulated true exposure value. The mutational contexts with 1 (left), 2 (middle) and 3 (right) flanking nucleotides on each side are shown for the additive model without opportunities (top row), additive model with opportunities (second row), interaction model without opportunities (third row) and interaction model with opportunities (bottom row). The identity line is displayed in gray for each panel.}
    \label{fig:accuracyW_BRCA}
\end{figure}
We consider a scenario inspired by Section \ref{subsec:breastres} with 4 signatures and 120 patients. Opportunities and signatures are fixed to those in Section \ref{subsec:breastres} for each of the models and context size. We simulate exposures from a Negative Binomial distribution with mean $\mu = 6000$ and dispersion $\alpha = 1.5$ following \cite{Pelizzola2023} and obtain the data $V$ for each scenario as $V_{nm} \sim \text{NB}\left( 200, \frac{(WHO)_{nm}}{200 + (WHO)_{nm}}\right)$.

We estimate the exposure and signature matrices with NMF with the additive and interaction models with and without opportunities to further demonstrate the effect of including opportunities in the model.

Figure \ref{fig:accuracyW_BRCA} shows the accuracy of estimating the exposure matrix in this analysis for the Tri-, Penta- and Hepta-nucleotide contexts for the  additive and interaction models with and without opportunities. While opportunities improve fit across all contexts, the Hepta-nucleotide results reveal the limit of the interaction model in this scenario; as the context size increases, the model becomes susceptible to overparametrization. This is resolved by the additive parametrization as shown in the two top panels of Figure \ref{fig:accuracyW_BRCA}, since this model further constrains the number of mutation types for the Hepta-nucleotide context. The results reported in this section show that the inclusion of opportunities is crucial for an accurate inference of the exposure matrix, in particular in extended contexts size. With these results, we further demonstrate that for a fixed model accuracy gains in exposure estimation are driven both by the extension to large context size and the inclusion of opportunities, provided the model is not overparametrized. 

\subsection{Results on the liver cancer data set} \label{subsec:liverres}
We consider here a data set with 260 liver cancer patients from the PCAWG database. Similar conclusions can be drawn also for this dataset. 
\begin{figure}[h]
    \centering
    \includegraphics[width =\textwidth]{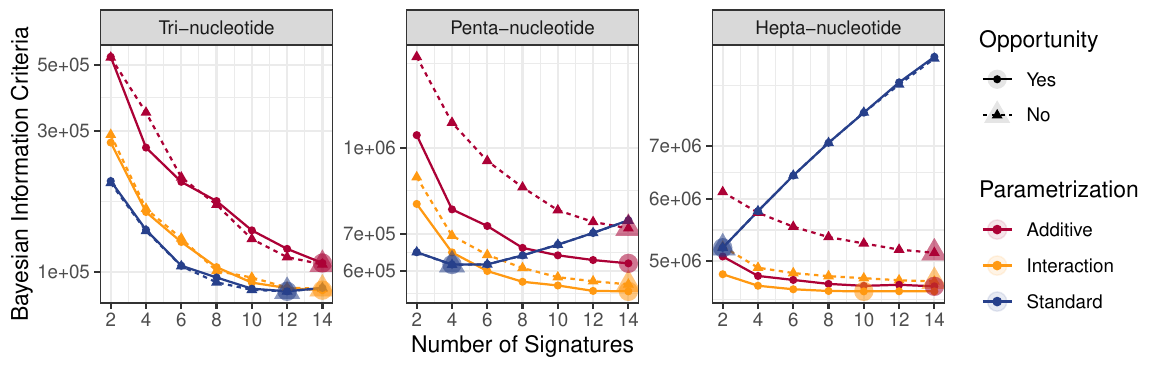}
    \caption{The influence of parametrizing and including opportunities on the estimate of the mutational counts in the liver data set. The BIC in log-scale plotted against the number of signatures. The minimum is highlighted by a larger point.}
    \label{fig:prediction_liver}
\end{figure}

Figure \ref{fig:prediction_liver} shows results on the influence of parametrizing and including the opportunities in estimating the original data where the BIC is plotted as a function of the number of signatures. With these results we show that parametrizing the signatures is a satisfactory compromise between decreasing the number of parameters and retaining good performance of the method. Decreasing the number of parameters is essential when looking at the Hepta-nucleotide context and in this context including the opportunities provides again a better fit, and thus lower log-likelihood values, in the parametrized models.

\section{Discussion} \label{discussion}
 
The mutation rate in a genome is strongly influenced by its local sequence context and the frequency of each context in the genome. This is referred to as the “mutational opportunity” and is often overlooked when analyzing mutational count data. Accounting for mutational opportunities is required for understanding mutational patterns, as it provides deeper insights into the underlying biological processes generating the observed mutations. In this work, we propose an extension of non-negative matrix factorization designed to extract robust mutational signatures in extended sequence contexts, and we show that correcting for the heterogeneity of the genome by accounting for mutational opportunities becomes essential to robust signature estimation in such contexts.

While classical NMF approaches and large contexts have been widely used, they typically assume uniform mutational opportunities across the genome. By explicitly incorporating extended sequence contexts and their corresponding mutational opportunities into the Negative Binomial NMF model, our approach enables more accurate estimation of both exposures and signatures while accounting for the heterogeneous values of opportunities across extended sequence contexts. To further improve stability and reduce overfitting, we integrate the parametrization technique introduced by \citep{laursen2024}, which minimizes the number of parameters needed to describe the signatures. This combined model demonstrates the importance of including mutational opportunities and parametrization for reliable exposures and signature estimation compared to traditional NMF in extended nucleotide contexts and with medium to large mutational opportunities. These model extensions consistently improve the estimation of exposures compared to both the tri-nucleotide model and formulations that do not account for mutational opportunities. In the setting examined in Section \ref{subsec:Waccuracy}, we find that models incorporating opportunities outperform those that do not across all configurations considered. Furthermore, based on these results, we recommend using the interaction parametrization for the signatures in small and intermediate sequence contexts, whereas the additive model gives better performance for the large hepta-nucleotide context. The latter observation is explained by the increased number of parameters required for the interaction model in larger contexts. In the cohort considered here, this results in reduced accuracy in estimating the exposures. This effect is expected to be less pronounced in larger cohorts, where the interaction model would be less affected by overparametrization in the hepta-nucleotide context and may become preferable in all contexts sizes. Taken together, these results indicate that the relative performance of the parametrizations is governed by the balance between model complexity and available data, while opportunities consistently improve estimation accuracy.

Considering extended sequence contexts and incorporating mutational opportunities facilitates the investigation of factors contributing to the overrepresentation of specific mutation types, such as DNA damage and repair mechanisms \citep{Poulsgaard2023}. By accurately quantifying overrepresented mutations, our approach allows for a better biological interpretability of signatures and a more direct way to link mutational patterns to the specific processes underlying mutagenesis. The inclusion of mutational opportunities is crucial for estimating signatures from indels data, as deletions and insertions often occur in repetitive genomic regions \citep{Streisinger1966}. Furthermore, tissue-specific analysis \citep{Cagan2022} would also benefit from the inclusion of opportunities for more interpretable results on tissue-specific mutagenesis, especially if multiple species are included. 

Developing models that yield robust signatures is essential for predicting accurate exposure for new data and for interpreting signatures correctly. By integrating mutational opportunities and applying parametrization, our approach represents a new methodology for extracting more stable mutational signatures.

\section*{Acknowledgement}
MP and AH acknowledges funding of the Novo Nordisk Foundation (Grant number NNF21OC0069105). Some of the computing for this project was performed on the GenomeDK cluster. We would like to thank GenomeDK and Aarhus University for providing computational resources.
\begin{appendices}

\section{Multiplicative update rules for NB-NMF with opportunities} \label{app:A}
For the MM algorithm, we construct a majorizing function $G(H, H^t)$ for  \\ $d_{\text{NBO}}(V||WH)$ with the constraint that $G(H, H) = d_{\text{NBO}}(V||WH)$.  Using Jensen's inequality and replacing $\log \left(\frac{\alpha_n + V_{nm}}{\alpha_n + \sum_{k=1}^K W_{nk}H_{km}O_m} \right)$ with the tangent line in $H^t$  because of the concavity property of the logarithm we obtain:

\begin{align}\label{eq:ghht}
\begin{split}
    d_{\text{NBO}}(V||WH) & = \sum_{n=1}^N \sum_{m=1}^M \left\{ V_{nm} \log \left( { \textstyle \frac{V_{nm}}{\sum_{k=1}^K W_{nk}H_{km}O_m} }\right) \right. \\ 
    & \hspace{1.8cm} \left.  - (\alpha_n + V_{nm}) \log \left({ \textstyle \frac{\alpha_n + V_{nm}}{\alpha_n + \sum_{k=1}^K W_{nk}H_{km}O_m} }\right) \right\} \\ 
    & \leq \sum_{n=1}^N \sum_{m=1}^M \left\{ V_{nm}  \log V_{nm} - V_{nm} \sum_{k=1}^K \gamma_{k} \log \left( { \textstyle \frac{W_{nk}H_{km}O_m}{\gamma_{k}}} \right) \right. \\ 
     & \hspace{1.8cm}   +  (\alpha_n + V_{nm})  \bigg[\log \left( { \textstyle \frac{\alpha_n + (WH^t)_{nm}O_m}{\alpha_n + V_{nm}} } \right) \\
     & \hspace{3cm} + \left. { \textstyle \frac{W_{nk}}{\alpha_n + (WH^t)_{nm}O_m} }(H_{km} - H^t_{km})O_m \bigg ] \right\} \\
    &  = G(H, H^t). 
\end{split}
\end{align} 
where $\gamma_{k} = \nicefrac{W_{nk}H_{km}^tO_m}{\sum_{k=1}^K W_{nk}H^t_{km}O_m}$. 
Lastly, it can easily be shown that $G(H, H) = d_{\text{NBO}}(V||WH)$ as follows:
\begin{align}
    \label{eq:ghh}
    \begin{split}
   G(H, H)& = \sum_{n=1}^N \sum_{m=1}^M \bigg\{ V_{nm} \log V_{nm}  \\
   & \hspace{1.8cm}  - V_{nm}\sum_{k=1}^K { \textstyle \frac{W_{nk}H_{km}O_m}{\sum_{k=1}^K W_{nk}H_{km}O_m} }\log \left ( { \textstyle \frac{W_{nk}H_{km}O_m}{  \left( {\scriptstyle \frac{W_{nk}H_{km}O_m}{\sum_{k=1}^K W_{nk}H_{km}O_m}} \right)}} \right)  \\ 
    &  \hspace{1.8cm}   - (\alpha_n + V_{nm}) \log \left( { \textstyle \frac{\alpha_n + V_{nm}}{\alpha_n + \sum_{k=1}^K W_{nk}H_{km}O_m} } \right) \bigg\} \\
    & =\sum_{n=1}^N  \sum_{m=1}^M \bigg\{ V_{nm} \log \left( { \textstyle \frac{V_{nm}}{\sum_{k=1}^K W_{nk}H_{km}O_m}} \right)  \\ 
    &  \hspace{1.8cm} - (\alpha_n + V_{nm}) \log \left( { \textstyle \frac{\alpha_n + V_{nm}}{\alpha_n + \sum_{k=1}^K W_{nk}H_{km}O_m}} \right) \bigg\} \\
    &= d_{\text{NBO}}(V||WH) 
    \end{split}
\end{align} 
Having defined the majorizing function $G(H, H^t)$ in \eqref{eq:ghht} we can derive the multiplicative updates for $H_{km}$ and $W_{nk}$ which leads to the update rules in \eqref{eq:updateH_methods} and \eqref{eq:updateW_methods}.

\newpage

\section{Parametric NMF with opportunities} \label{app:B}

\begin{algorithm}[!h] 
\SetAlgoLined
Given data matrix $V$, rank $K$, design matrices $X_1, \dots , X_K$, opportunity vector $O = (O_1,\dots, O_M)$ and threshold $\epsilon$.\\ \vspace{0.2cm}
Estimate dispersion parameters: \\
\quad Get $W^{Po}, H^{Po}$ by applying poisson NMF updates to $V$ with $K$ signatures \\
\quad Estimate $\alpha_1, \dots , \alpha_N$ by Negative Binomial MLE using $W^{Po}, H^{Po}$ and $V$ \\ \vspace{0.2cm}

Initialize $W^1$ and $H^1$ with random entries. \\
 \For{ $t = 1,2,3, \dots $ }{
 \For{ $k = 1,\dots, K$}{
 Update each signature
       $$H_k^t = H_k^t \otimes \frac{\left( (W_k^t)' \frac{V}{W^t H^t} \right) \frac{1}{O}}{\left( (W_k^t)' \frac{V + \alpha_n}{W^t H^t O + \alpha_n} \right)}$$
      Fit the log-linear Poisson regression \\
      \begin{equation}
    \log ( H_k^t) = X_k \beta_k^t
\end{equation}
      for estimating $\beta_k^t$ and set 
     $$H_k^{t+1}(\hat{\beta}_k^t) = \frac{ \exp(X_k \hat{\beta}_k^t) }{ \textbf{1}' \exp(X_k \hat{\beta}_k^t) } $$\\
 }
 Update exposures
\begin{equation*}
     W^{t+1} = W^t \otimes \frac{\left( \frac{V}{W^t H^t(\hat{\beta}^t)} (H^t(\hat{\beta}^t))' \right)}{\left( \frac{V + \alpha_n}{W^t H^t(\hat{\beta}^t) O + \alpha_n} (H^t(\hat{\beta}^t))' \right)\frac{1}{O}}
\end{equation*}
\textbf{stop if} $\frac{\ell(W^{t+1},H^{t+1}(\hat{\beta}^{t});Z)-\ell(W^t,H^t(\hat{\beta}^{t-1});Z)}{\ell(W^{t+1},H^{t+1}(\hat{\beta}^{t});Z)} < \epsilon$
 }
\caption{Parametric NMF with opportunities.}
\label{alg:NMFopp}
\end{algorithm}

\newpage
\section{Residual plots as a function of the opportunities} \label{app:C}
In this Section we show the standardized residuals as a function of the opportunity percentiles for the interaction model with (right) and without (left) opportunities for the Tri- (top row) Penta- (middle row) and Hepta-nucleotide context (bottom row).

\begin{figure}[h!]
    \centering
    \includegraphics[width = \textwidth]{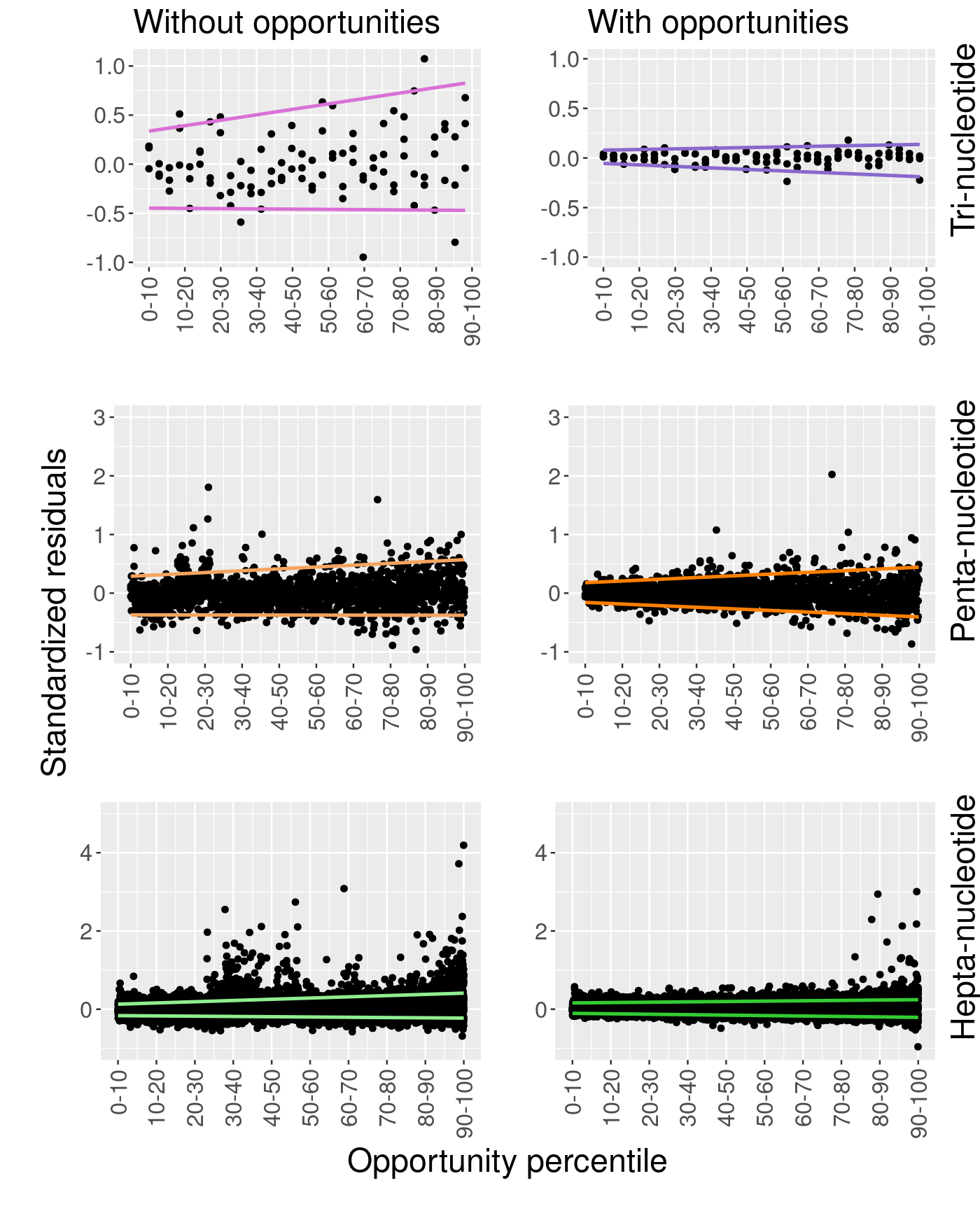}
    \caption{The standardized residuals for the interaction model with and without opportunities as a function of the opportunity percentile.}
    \label{fig:res_vs_opp}
\end{figure}

\end{appendices}

\newpage 

\bibliographystyle{apalike}
\bibliography{bibliography}

@article{kmerpapa2022,
  doi = {10.1038/s41467-022-35596-5},
  url = {https://doi.org/10.1038/s41467-022-35596-5},
  year = {2022},
  month = dec,
  publisher = {Springer Science and Business Media {LLC}},
  volume = {13},
  number = {1},
  author = {J\"{o}rn Bethune and April Kleppe and S{\o}ren Besenbacher},
  title = {A method to build extended sequence context models of point mutations and indels},
  journal = {Nature Communications}
}

@article{laursen2024,
  title={Flexible model-based non-negative matrix factorization with application to mutational signatures},
  author={Laursen, Ragnhild and Maretty, Lasse and Hobolth, Asger},
  journal={Statistical Applications in Genetics and Molecular Biology},
  volume={23},
  number={1},
  pages={20230034},
  year={2024},
  publisher={De Gruyter},
  url = {https://doi.org/10.1515/sagmb-2023-0034}
}

@article{Pelizzola2023,
  doi = {10.1186/s12859-023-05304-1},
  url = {https://doi.org/10.1186/s12859-023-05304-1},
  year = {2023},
  month = may,
  publisher = {Springer Science and Business Media {LLC}},
  volume = {24},
  number = {1},
  author = {Marta Pelizzola and Ragnhild Laursen and Asger Hobolth},
  title = {Model selection and robust inference of mutational signatures using Negative Binomial non-negative matrix factorization},
  journal = {{BMC} Bioinformatics}
}

@article{alexandrov2013signatures,
  title={Signatures of mutational processes in human cancer},
  author={Alexandrov, Ludmil B and Nik-Zainal, Serena and Wedge, David C and Aparicio, Samuel AJR and Behjati, Sam and Biankin, Andrew V and Bignell, Graham R and Bolli, Niccolo and Borg, Ake and B{\o}rresen-Dale, Anne-Lise and others},
  journal={Nature},
  volume={500},
  number={7463},
  pages={415--421},
  year={2013},
  publisher={Nature Publishing Group},
  url = {https://doi.org/10.1038/nature12477}
}

@article{alexandrov2013breastcancer,
      title={Deciphering signatures of mutational processes operative in human cancer.}, 
      author={Alexandrov, L. B. and Nik-Zainal, S. and Wedge, D. C.  and Campbell, P. J. and Stratton, M. R.},
      year={2013},
      journal={Cell reports},
      number = {1},
      volume = {3},
      pages = {264--259},
      url = {https://doi.org/10.1016/j.celrep.2012.12.008}
}

@article{Gouvert2020,
    author = {Gouvert, Olivier and Oberlin, Thomas and Fevotte, Cedric},
    doi = {10.1109/LSP.2020.2991613},
    issn = {15582361},
    journal = {IEEE Signal Processing Letters},
    pages = {815--819},
    publisher = {Institute of Electrical and Electronics Engineers Inc.},
    title = {{Negative Binomial Matrix Factorization}},
    volume = {27},
    year = {2020},
    url = {https://hal.science/hal-02871905v1}
}

@article{Fischer2013,
    author = {Fischer, Andrej and Illingworth, Christopher J.R. and Campbell, Peter J. and Mustonen, Ville},
    doi = {10.1186/gb-2013-14-4-r39},
    issn = {1474760X},
    journal = {Genome Biology},
    month = {apr},
    number = {4},
    pages = {1--10},
    pmid = {23628380},
    publisher = {BioMed Central},
    title = {{EMu: Probabilistic inference of mutational processes and their localization in the cancer genome}},
    volume = {14},
    year = {2013},
    url = {https://doi.org/10.1186/gb-2013-14-4-r39}
}

@article{Lal2021,
    author = {Lal, Avantika and Liu, Keli and Tibshirani, Robert and Sidow, Arend and Ramazzotti, Daniele},
    doi = {10.1371/JOURNAL.PCBI.1009119},
    issn = {1553-7358},
    journal = {PLOS Computational Biology},
    month = {jun},
    number = {6},
    pages = {e1009119},
    publisher = {Public Library of Science},
    title = {{De novo mutational signature discovery in tumor genomes using SparseSignatures}},
    volume = {17},
    year = {2021},
    url = {https://doi.org/10.1371/journal.pcbi.1009119}
}

@article{Lee1999,
    author = {Lee, Daniel D. and Seung, H. Sebastian},
    doi = {10.1038/44565},
    issn = {00280836},
    journal = {Nature},
    month = {oct},
    number = {6755},
    pages = {788--791},
    pmid = {10548103},
    publisher = {Nature Publishing Group},
    title = {{Learning the parts of objects by non-negative matrix factorization}},
    volume = {401},
    year = {1999},
    url = {https://doi.org/10.1038/44565}
}

@article{Lyu2020,
    author = {Lyu, Xinrui and Garret, Jean and R{\"{a}}tsch, Gunnar and Lehmann, Kjong Van},
    doi = {10.1093/BIOINFORMATICS/BTAA473},
    issn = {14602059},
    journal = {Bioinformatics},
    month = {jul},
    number = {Suppl{\_}1},
    pages = {i154--i160},
    pmid = {32657388},
    publisher = {Oxford University Press},
    title = {{Mutational signature learning with supervised negative binomial non-negative matrix factorization}},
    volume = {36},
    year = {2020},
    url = {https://doi.org/10.1093/bioinformatics/btaa473}
}

@article{Campbell2020,
  title = {Pan-cancer analysis of whole genomes},
  volume = {578},
  ISSN = {1476-4687},
  url = {http://dx.doi.org/10.1038/s41586-020-1969-6},
  DOI = {10.1038/s41586-020-1969-6},
  number = {7793},
  journal = {Nature},
  publisher = {Springer Science and Business Media LLC},
  author = {{The ICGC/TCGA Pan-Cancer Analysis of Whole Genomes Consortium}},
  year = {2020},
  month = feb,
  pages = {82–93}
}

@article{Weinhold2014,
    author = {Weinhold, N and Jacobsen, A and Schultz, N and Sander, C and Lee, W},
    doi = {10.1038/NG.3101},
    issn = {1546-1718},
    journal = {Nature genetics},
    month = {nov},
    number = {11},
    pages = {1160--1165},
    pmid = {25261935},
    publisher = {Nat Genet},
    title = {{Genome-wide analysis of noncoding regulatory mutations in cancer}},
    volume = {46},
    year = {2014},
    url = {https://doi.org/10.1038/ng.3101}
}

@article{Lochovsky2015,
    author = {Lochovsky, L and Zhang, J and Fu, Y and Khurana, E and Gerstein,  M},
    doi = {10.1093/NAR/GKV803},
    issn = {1362-4962},
    journal = {Nucleic acids research},
    month = {sep},
    number = {17},
    pages = {8123--8134},
    pmid = {26304545},
    publisher = {Nucleic Acids Res},
    title = {{LARVA: an integrative framework for large-scale analysis of recurrent variants in noncoding annotations}},
    volume = {43},
    year = {2015},
    url = {https://doi.org/10.1093/nar/gkv803}
}

@article{Risques2018,
    author = {Risques, Rosa Ana and Kennedy, Scott R.},
    doi = {10.1371/JOURNAL.PGEN.1007108},
    journal = {PLOS Genetics},
    month = {jan},
    number = {1},
    pmid = {29300727},
    publisher = {Public Library of Science},
    title = {{Aging and the rise of somatic cancer-associated mutations in normal tissues}},
    volume = {14},
    year = {2018},
    url = {https://doi.org/10.1371/journal.pgen.1007108}
}

@article{Alexandrov2016,
    author = {Alexandrov, Ludmil B. and Ju, Young Seok and Haase, Kerstin and {Van Loo}, Peter and Martincorena, I{\~{n}}igo and Nik-Zainal, Serena and Totoki, Yasushi and Fujimoto, Akihiro and Nakagawa, Hidewaki and Shibata, Tatsuhiro and Campbell, Peter J. and Vineis, Paolo and Phillips, David H. and Stratton, Michael R.},
    url = {https://doi.org/10.1126/science.aag0299},
    journal = {Science},
    month = {nov},
    number = {6312},
    pages = {618--622},
    publisher = {American Association for the Advancement of Science},
    title = {{Mutational signatures associated with tobacco smoking in human cancer}},
    volume = {354},
    year = {2016}
}

@article{Shibai2017,
    author = {Shibai, Atsushi and Takahashi, Yusuke and Ishizawa, Yuka and Motooka, Daisuke and Nakamura, Shota and Ying, Bei-Wen and Tsuru, Saburo},
    issn = {2045-2322},
    journal = {Scientific Reports},
    month = {nov},
    number = {1},
    pages = {1--12},
    publisher = {Nature Publishing Group},
    title = {{Mutation accumulation under UV radiation in Escherichia coli}},
    volume = {7},
    year = {2017},
    url = {https://doi.org/10.1038/s41598-017-15008-1}
}

@article{Alexandrov2020,
    author = {Alexandrov, Ludmil B. and Kim, Jaegil and Haradhvala, Nicholas J. and Huang, Mi Ni and {Tian Ng}, Alvin Wei and Wu, Yang and Boot, Arnoud and Covington, Kyle R. and Gordenin, Dmitry A. and Bergstrom, Erik N. and Islam, S. M. Ashiqul and Lopez-Bigas, Nuria and Klimczak, Leszek J. and McPherson, John R. and Morganella, Sandro and Sabarinathan, Radhakrishnan and Wheeler, David A. and Mustonen, Ville and Getz, Gad and Rozen, Steven G. and Stratton, Michael R.},
    doi = {10.1038/s41586-020-1943-3},
    issn = {1476-4687},
    journal = {Nature},
    month = {feb},
    number = {7793},
    pages = {94--101},
    publisher = {Nature Publishing Group},
    title = {{The repertoire of mutational signatures in human cancer}},
    volume = {578},
    year = {2020}, 
    url = {https://doi.org/10.1038/s41586-020-1943-3}
}

@article{Tate2019,
    author = {Tate, John G and Bamford, Sally and Jubb, Harry C and Sondka, Zbyslaw and Beare, David M and Bindal, Nidhi and Boutselakis, Harry and Cole, Charlotte G and Creatore, Celestino and Dawson, Elisabeth and Fish, Peter and Harsha, Bhavana and Hathaway, Charlie and Jupe, Steve C and Kok, Chai Yin and Noble, Kate and Ponting, Laura and Ramshaw, Christopher C and Rye, Claire E and Speedy, Helen E and Stefancsik, Ray and Thompson, Sam L and Wang, Shicai and Ward, Sari and Campbell, Peter J and Forbes, Simon A},
    doi = {10.1093/NAR/GKY1015},
    issn = {0305-1048},
    journal = {Nucleic Acids Research},
    month = {jan},
    number = {D1},
    pages = {D941--D947},
    publisher = {Oxford Academic},
    title = {{COSMIC: the Catalogue Of Somatic Mutations In Cancer}},
    volume = {47},
    year = {2019},
    url = {https://doi.org/10.1093/nar/gky1015}
}

@article {Gori2018,
	author = {Gori, Kevin and Baez-Ortega, Adrian},
	title = {sigfit: flexible Bayesian inference of mutational signatures},
	elocation-id = {372896},
	year = {2020},
	volume = {https://doi.org/10.1101/372896},
	publisher = {Cold Spring Harbor Laboratory},
	URL = {https://www.biorxiv.org/content/early/2020/01/17/372896},
	eprint = {https://www.biorxiv.org/content/early/2020/01/17/372896.full.pdf},
	journal = {bioRxiv}
}

@article{Vhringer2021,
  doi = {10.1038/s41467-021-23551-9},
  url = {https://doi.org/10.1038/s41467-021-23551-9},
  year = {2021},
  month = jun,
  publisher = {Springer Science and Business Media {LLC}},
  volume = {12},
  number = {1},
  author = {Harald V\"{o}hringer and Arne Van Hoeck and Edwin Cuppen and Moritz Gerstung},
  title = {Learning mutational signatures and their multidimensional genomic properties with {TensorSignatures}},
  journal = {Nature Communications}
}

@article{Dietlein2020,
  doi = {10.1038/s41588-019-0572-y},
  url = {https://doi.org/10.1038/s41588-019-0572-y},
  year = {2020},
  month = feb,
  publisher = {Springer Science and Business Media {LLC}},
  volume = {52},
  number = {2},
  pages = {208--218},
  author = {Felix Dietlein and Donate Weghorn and Amaro Taylor-Weiner and Andr{\'{e}} Richters and Brendan Reardon and David Liu and Eric S. Lander and Eliezer M. Van Allen and Shamil R. Sunyaev},
  title = {Identification of cancer driver genes based on nucleotide context},
  journal = {Nature Genetics}
}

@article{Omichessan2019,
  doi = {10.1371/journal.pone.0221235},
  url = {https://doi.org/10.1371/journal.pone.0221235},
  year = {2019},
  month = sep,
  publisher = {Public Library of Science ({PLoS})},
  volume = {14},
  number = {9},
  pages = {e0221235},
  author = {Hanane Omichessan and Gianluca Severi and Vittorio Perduca},
  editor = {Alvaro Galli},
  title = {Computational tools to detect signatures of mutational processes in {DNA} from tumours: A review and empirical comparison of performance},
  journal = {{PLOS} {ONE}}
}

@article{BaezOrtega2017,
  doi = {10.1093/bib/bbx082},
  url = {https://doi.org/10.1093/bib/bbx082},
  year = {2017},
  month = jul,
  publisher = {Oxford University Press ({OUP})},
  volume = {20},
  number = {1},
  pages = {77--88},
  author = {Adrian Baez-Ortega and Kevin Gori},
  title = {Computational approaches for discovery of mutational signatures in cancer},
  journal = {Briefings in Bioinformatics}
}

@article{Lindberg2019,
  doi = {10.1073/pnas.1909021116},
  url = {https://doi.org/10.1073/pnas.1909021116},
  year = {2019},
  month = sep,
  publisher = {Proceedings of the National Academy of Sciences U.S.A.},
  volume = {116},
  number = {41},
  pages = {20411--20417},
  author = {Markus Lindberg and Martin Bostr\"{o}m and Kerryn Elliott and Erik Larsson},
  title = {Intragenomic variability and extended sequence patterns in the mutational signature of ultraviolet light},
  journal = {Proceedings of the National Academy of Sciences}
}

@article{Shiraishi2015,
  doi = {10.1371/journal.pgen.1005657},
  url = {https://doi.org/10.1371/journal.pgen.1005657},
  year = {2015},
  month = dec,
  publisher = {Public Library of Science ({PLoS})},
  volume = {11},
  number = {12},
  pages = {e1005657},
  author = {Yuichi Shiraishi and Georg Tremmel and Satoru Miyano and Matthew Stephens},
  editor = {Jonathan Marchini},
  title = {A Simple Model-Based Approach to Inferring and Visualizing Cancer Mutation Signatures},
  journal = {{PLOS} Genetics}
}

@article{Caruso2017,
  title = {Niraparib in ovarian cancer: results to date and clinical potential},
  volume = {9},
  ISSN = {1758-8359},
  url = {http://dx.doi.org/10.1177/1758834017718775},
  DOI = {10.1177/1758834017718775},
  number = {9},
  journal = {Therapeutic Advances in Medical Oncology},
  publisher = {SAGE Publications},
  author = {Caruso,  Davide and Papa,  Anselmo and Tomao,  Silverio and Vici,  Patrizia and Panici,  Pierluigi Benedetti and Tomao,  Federica},
  year = {2017},
  month = jul,
  pages = {579–588}
}

@article{Zhang2021,
  title = {Genomic and evolutionary classification of lung cancer in never smokers},
  volume = {53},
  ISSN = {1546-1718},
  url = {http://dx.doi.org/10.1038/s41588-021-00920-0},
  DOI = {10.1038/s41588-021-00920-0},
  number = {9},
  journal = {Nature Genetics},
  publisher = {Springer Science and Business Media LLC},
  author = {Zhang,  Tongwu and Joubert,  Philippe and Ansari-Pour,  Naser and Zhao,  Wei and Hoang,  Phuc H. and Lokanga,  Rachel and Moye,  Aaron L. and Rosenbaum,  Jennifer and Gonzalez-Perez,  Abel and Martínez-Jiménez,  Francisco and Castro,  Andrea and Muscarella,  Lucia Anna and Hofman,  Paul and Consonni,  Dario and Pesatori,  Angela C. and Kebede,  Michael and Li,  Mengying and Gould Rothberg,  Bonnie E. and Peneva,  Iliana and Schabath,  Matthew B. and Poeta,  Maria Luana and Costantini,  Manuela and Hirsch,  Daniela and Heselmeyer-Haddad,  Kerstin and Hutchinson,  Amy and Olanich,  Mary and Lawrence,  Scott M. and Lenz,  Petra and Duggan,  Maire and Bhawsar,  Praphulla M. S. and Sang,  Jian and Kim,  Jung and Mendoza,  Laura and Saini,  Natalie and Klimczak,  Leszek J. and Islam,  S. M. Ashiqul and Otlu,  Burcak and Khandekar,  Azhar and Cole,  Nathan and Stewart,  Douglas R. and Choi,  Jiyeon and Brown,  Kevin M. and Caporaso,  Neil E. and Wilson,  Samuel H. and Pommier,  Yves and Lan,  Qing and Rothman,  Nathaniel and Almeida,  Jonas S. and Carter,  Hannah and Ried,  Thomas and Kim,  Carla F. and Lopez-Bigas,  Nuria and Garcia-Closas,  Montserrat and Shi,  Jianxin and Bossé,  Yohan and Zhu,  Bin and Gordenin,  Dmitry A. and Alexandrov,  Ludmil B. and Chanock,  Stephen J. and Wedge,  David C. and Landi,  Maria Teresa},
  year = {2021},
  month = sep,
  pages = {1348–1359}
}

@article{Streisinger1966,
  title = {Frameshift Mutations and the Genetic Code},
  volume = {31},
  ISSN = {1943-4456},
  url = {http://dx.doi.org/10.1101/SQB.1966.031.01.014},
  DOI = {10.1101/sqb.1966.031.01.014},
  number = {0},
  journal = {Cold Spring Harbor Symposia on Quantitative Biology},
  publisher = {Cold Spring Harbor Laboratory},
  author = {Streisinger,  G. and Okada,  Y. and Emrich,  J. and Newton,  J. and Tsugita,  A. and Terzaghi,  E. and Inouye,  M.},
  year = {1966},
  month = jan,
  pages = {77–84}
}

@article{Cagan2022,
  title = {Somatic mutation rates scale with lifespan across mammals},
  volume = {604},
  ISSN = {1476-4687},
  url = {http://dx.doi.org/10.1038/s41586-022-04618-z},
  DOI = {10.1038/s41586-022-04618-z},
  number = {7906},
  journal = {Nature},
  publisher = {Springer Science and Business Media LLC},
  author = {Cagan,  Alex and Baez-Ortega,  Adrian and Brzozowska,  Natalia and Abascal,  Federico and Coorens,  Tim H. H. and Sanders,  Mathijs A. and Lawson,  Andrew R. J. and Harvey,  Luke M. R. and Bhosle,  Shriram and Jones,  David and Alcantara,  Raul E. and Butler,  Timothy M. and Hooks,  Yvette and Roberts,  Kirsty and Anderson,  Elizabeth and Lunn,  Sharna and Flach,  Edmund and Spiro,  Simon and Januszczak,  Inez and Wrigglesworth,  Ethan and Jenkins,  Hannah and Dallas,  Tilly and Masters,  Nic and Perkins,  Matthew W. and Deaville,  Robert and Druce,  Megan and Bogeska,  Ruzhica and Milsom,  Michael D. and Neumann,  Bj\"{o}rn and Gorman,  Frank and Constantino-Casas,  Fernando and Peachey,  Laura and Bochynska,  Diana and Smith,  Ewan St. John and Gerstung,  Moritz and Campbell,  Peter J. and Murchison,  Elizabeth P. and Stratton,  Michael R. and Martincorena,  Iñigo},
  year = {2022},
  month = apr,
  pages = {517–524}
}

@article{Poulsgaard2023,
  title = {Sequence dependencies and mutation rates of localized mutational processes in cancer},
  volume = {15},
  ISSN = {1756-994X},
  url = {http://dx.doi.org/10.1186/s13073-023-01217-z},
  DOI = {10.1186/s13073-023-01217-z},
  number = {1},
  journal = {Genome Medicine},
  publisher = {Springer Science and Business Media LLC},
  author = {Poulsgaard,  Gustav Alexander and Sørensen,  Simon Grund and Juul,  Randi Istrup and Nielsen,  Morten Muhlig and Pedersen,  Jakob Skou},
  year = {2023},
  month = aug 
}

\end{document}